\documentclass[journal=ancac3,manuscript=article]{achemso}

\usepackage{graphicx}
\usepackage[version=3]{mhchem} 
\usepackage{epstopdf}
\usepackage[usenames,dvipsnames]{xcolor}
\usepackage{physics}
\usepackage{longtable}
\usepackage{multirow}

\newcommand{\Fig}[1]{Fig.\ref{#1}}
\newcommand{\Figure}[1]{Figure~\ref{#1}}
\newcommand{\SI}[1] {Supporting Information}

\newcommand{\Tab}[1]{table~\ref{#1}}

\newcommand{\refone}[1]{\textcolor{black}{#1}}
\newcommand{\reftwo}[1]{\textcolor{black}{#1}}

\author{Rocco Gaudenzi}
\author{Joeri de Bruijckere}
\affiliation{Kavli Institute of Nanoscience, Delft University of Technology, Lorentzweg 1, 2628 CJ Delft, The Netherlands}
\author{Daniel Reta}
\affiliation{Departament de Ci\`encia de Materials i Qu\'imica F\'isica and Institut de Qu\'imica Te\'orica i Computacional, Universitat de Barcelona (IQTCUB), E-08028 Barcelona, Spain}
\alsoaffiliation{now at School of Chemistry, The University of Manchester, Oxford Road, Manchester, M13 9PL United Kingdom}
\author{Ib\'erio de P. R. Moreira}
\affiliation{Departament de Ci\`encia de Materials i Qu\'imica F\'isica and Institut de Qu\'imica Te\'orica i Computacional, Universitat de Barcelona (IQTCUB), E-08028 Barcelona, Spain}
\author{Concepci\'o Rovira}
\affiliation{Institut de Ci\`enca de Materials de Barcelona (ICMAB-CSIC) and CIBER-BBN, Campus de la UAB, 08193, Bellaterra, Spain}
\author{Jaume Veciana}
\affiliation{Institut de Ci\`enca de Materials de Barcelona (ICMAB-CSIC) and CIBER-BBN, Campus de la UAB, 08193, Bellaterra, Spain}
\author{Herre S. J. van der Zant}
\author{Enrique Burzur\'i}
\email{E.BurzuriLinares@tudelft.nl}
\affiliation{Kavli Institute of Nanoscience, Delft University of Technology, Lorentzweg 1, 2628 CJ Delft, The Netherlands}

\date{\today}

\title{Redox-Induced Gating of the Exchange Interactions in a Single Organic Diradical}


\begin{document}
\clearpage
\begin{abstract}

Embedding a magnetic electroactive molecule in a three-terminal junction allows for the fast and local electric field control of magnetic properties desirable in spintronic devices and quantum gates. 
Here, we provide an example of this control through the reversible and stable charging of a single all-organic \refone{neutral} diradical molecule.
By means of inelastic electron tunnel spectroscopy (IETS) we show that the added
electron occupies a molecular orbital distinct from those containing the two radical electrons, forming a spin system with three antiferromagnetically-coupled spins. 
\refone{Changing the redox state of the molecule therefore switches on and off a parallel exchange path between the two radical spins through the added electron.}
This electrically-controlled gating of the intramolecular magnetic interactions constitutes an essential ingredient of a single-molecule $\sqrt{\text{SWAP}}$ quantum gate. 
\end{abstract}


Fast, reversible and local control of magnetic properties of molecular systems is sought for as a potential path for 
molecule-based spintronic devices \cite{Wolf2001, Sanvito2011, Urdampilleta2011} and quantum information processing.\cite{Bogani2008, Vincent2012, Thiele2014} 
The control of the intramolecular exchange coupling could allow, for instance, for the realization of a single-molecule quantum gate.\cite{Lehmann2009, Troiani2011, Ferrando2016a, Ferrando2016b} 
One way to achieve such control at the single-molecule level is to embed a magnetic electro-active molecule in a solid-state junction and use the gate electrode to change its magnetic properties through a form of spin-electric coupling. \cite{Trif2008, Lehmann2009, Osorio2010, Trif2010, Islam2010, Florens2011, Palii2014, Cardona2015} Traditional candidates are single-molecule magnets (SMMs), the magnetic parameters of which can be modulated with the addition of a charge \cite{Burzuri2012a, Nossa2013, Misiorny2015} or through magnetoelectric effects. \cite{Scarrozza2016} 

A promising alternative \refone{to SMMs} is offered by all-organic radical molecules \cite{Lehmann2009} where the magnetism arises from the unpaired spins of carbon atoms.\cite{Rajca2001, li2016} The simplicity of their \reftwo{spin structure} and the absence of metal centers have proven to yield robust molecular junctions \cite{Mas2009, Frisenda2015, Gaudenzi2016} and potentially allow to overcome the limitations inherent to SMMs owing to low spin-orbit coupling and hyperfine interaction. However, the existing experimental examples have shown either a relatively small electric control over the exchange coupling \cite{Gaudenzi2016} or a reduction of the molecule to a closed-shell system with no unpaired spins.\cite{Simao2011} 

\begin{figure}
	\includegraphics{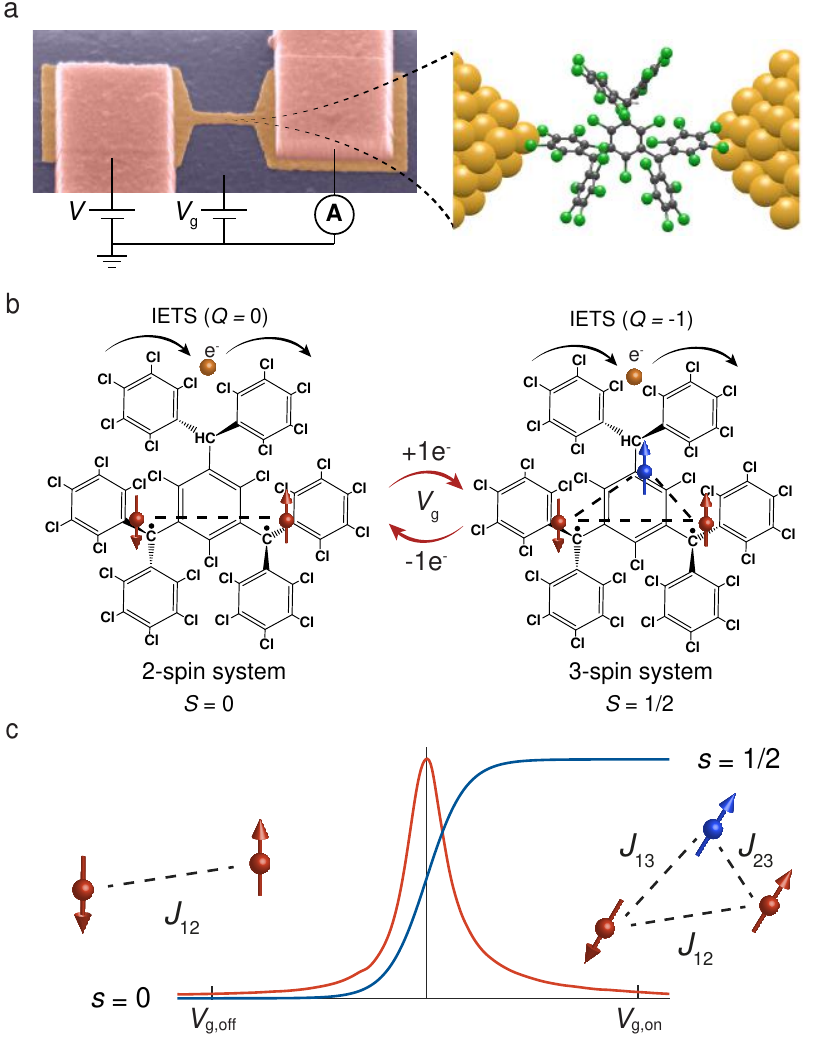}
	\caption{\textbf{The organic diradical spin system.} (a) Scanning Electron Microscope (SEM) false color image of a Au nanowire on top of an Al$_2$O$_3$/AuPd gate. (b) Structure and magnetism schematics of the neutral diradical and \refone{reduced} form of the diradical. The red dots and the dashed lines mark the radical spins and the exchange interactions, respectively. A gate voltage allows to reversibly add a spin (blue dot) onto the redox center and, with that, switch on and off the magnetic couplings between \refone{the added electron} and the two radical spins. For each state inelastic electron tunneling spectroscopy (IETS) is performed (yellow electron). (c) Differential conductance (red) and corresponding redox center spin value $s$ (blue) as a function of $V_\textrm{g}$. Sweeping from $V_\textrm{g,off}$  to $V_\textrm{g,on}$, the site is progressively filled and $s$ increases from 0 to 1/2. The value $V_\textrm{g,off}$ ($V_\textrm{g,on}$) marks the gate voltage at which the added spin stably resides off (on) the molecule.
	}
	\label{figure1}
\end{figure}

Here, we report the reversible and stable reduction of a \refone{neutral} diradical molecule in a three-terminal 
device, by means of a gate electrode. Inelastic electron tunneling spectroscopy (IETS) in the two stable redox states shows that the added electron magnetically couples to the two radical spins, preserving their open-shell character, while changing the magnetic state of the molecule from a singlet to a doublet state with three unpaired electrons. 
This ability to reversibly switch on the exchange couplings between the added electron and the two radical spins could form the base for a \mbox{$\sqrt{\text{SWAP}}$ quantum gate.\cite{Lehmann2007, Lehmann2009, Luis2011}} 

\section{RESULTS AND DISCUSSION}

The molecule we use is a neutral 2,4,6-hexakis(pentachlorophenyl)mesitylene diradical molecule,\cite{Veciana1993} hereafter PTM-based diradical, schematically shown in \Figure{figure1}. It is made of three methyl carbon atoms connected \textit{via} a central \refone{benzene} ring. Two of \refone{these C atoms are methyl radicals} with unpaired electrons, while the third binds \refone{a H that closes the electronic shell. The resulting molecule is a two-spin magnetic system.} Two chlorinated phenyl rings attach and surround each methyl carbon in a propeller-like configuration as seen in \Figure{figure1}. The single-molecule junction is formed when a single PTM-based diradical bridges the source and drain electrodes as illustrated in \Figure{figure1}(a). The electric field produced by applying a gate voltage $V_\textrm{g}$ to the third electrode is used to change the redox state of the molecule (\Figure{figure1}(b)). In transitioning between the two states, a high-conductance peak is traversed. On the right (left) of the peak, \textit{i.e.}, at $V_\textrm{g,on}$ ($V_\textrm{g,off}$), the redox center has a stable spin $s = 1/2$ ($s = 0$). Additional details on the fabrication and molecule deposition can be found elsewhere \cite{Burzuri2015, Perrin2015} and in Methods.

\begin{figure}
	\includegraphics{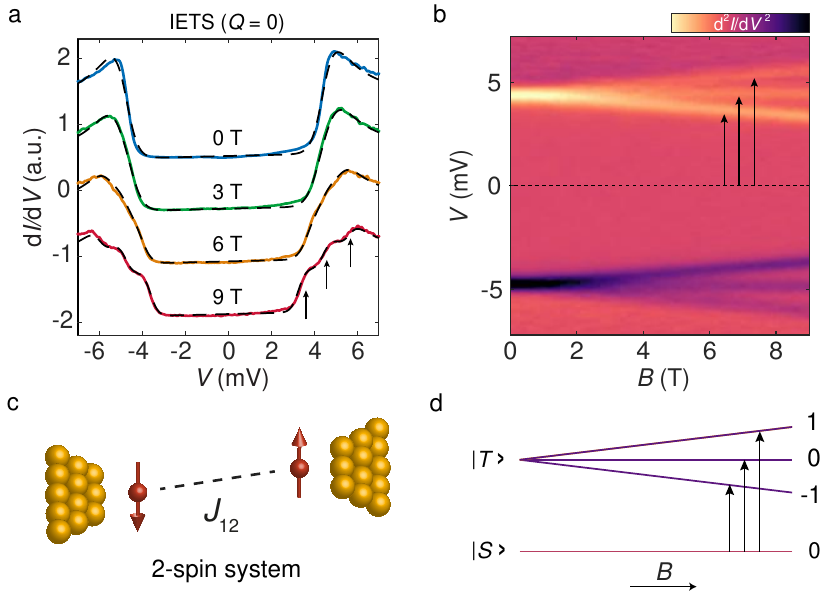}
	\caption{\textbf{Magnetic spectrum of the neutral diradical.} (a) Differential conductance (d$I$/d$V$) spectra of the \refone{neutral} diradical molecular junction at different magnetic fields and at a fixed gate voltage $V_\textrm{g}$ = -2.3 V. An excitation step at $\pm4.65$ mV splits in three substeps under applied magnetic field. The superimposed dashed lines are fits using the model in Ref. \cite{Ternes2015}. (b) d$^{2}I$/d$V^{2}$ color map showing the splitting as a function of $V$ and $B$. (c) Schematics of a two-spin system with exchange coupling $J_{12}$, confined between two gold electrodes. (d) Spin spectrum and allowed transitions for a two-spin system with antiferromagnetic $J_{12}$.}
	\label{figure2}
\end{figure}

We probe the excitation spectrum of an individual diradical molecule by measuring the \textit{dc}-current $I$ through the junction as a function of bias voltage $V$ and extracting the differential conductance d$I$/d$V$. Each step in the d$I$/d$V$ spectrum signals the opening of an inelastic electron current channel \textit{via} the excited state of the molecule with energy $eV$. Following the steps' energy as a function of magnetic field allows to read out the molecule's energy spectrum, providing a single-molecule analogue of electron-spin resonance spectroscopy. \Figure{figure2}(a) shows the d$I$/d$V$ spectra of a diradical junction at different magnetic fields $B$ at fixed gate voltage $V_\textrm{g} = -2.3 $ V. The spectrum taken at 0 T shows symmetric steps at $\pm 4.65$ mV, which can be associated with transitions to excited spin states. The confirmation of the magnetic nature of the transitions is given by the evolution of the excitation energies as a function of the applied magnetic field $B$ (see \Figure{figure2}(b), where the second derivative, $\textrm{d}^2I/\textrm{d}V^2$, is shown for clarity). The excitation step splits in three substeps as $B$ is increased. As shown in the level scheme of \Figure{figure2}(d), this spectrum is consistent with the anti-ferromagnetically coupled two-spin system depicted in \Figure{figure2}(c) with an open-shell singlet ($S$ = 0) ground state $\left|S\right>$ and a triplet ($S$ = 1) excited state $\left|T\right>$.

We compare the experimental spectra with numerical simulations based on the tunneling model of Ref. \cite{Ternes2015}, commonly used in scanning tunneling spectroscopy. The dashed black lines in \Figure{figure2}(a) show the results of these simulations. Within the framework of the model, we describe the diradical molecule by a model Hamiltonian with two spin-1/2 centers interacting through a Heisenberg exchange coupling $J_{12}$. For all magnetic field values, the data can be well fitted to this model with $J_{12}$ = 4.65 meV. 
The preference for the singlet ground state is ascribed to the distortion of the molecule in the solid state device in analogy with previous studies on PTM-based \refone{neutral} triradicals.\cite{Gaudenzi2016}
We have verified the plausibility of this scenario by DFT calculations (see \SI~ Section 2.2 for details).  

\begin{figure}
	\includegraphics{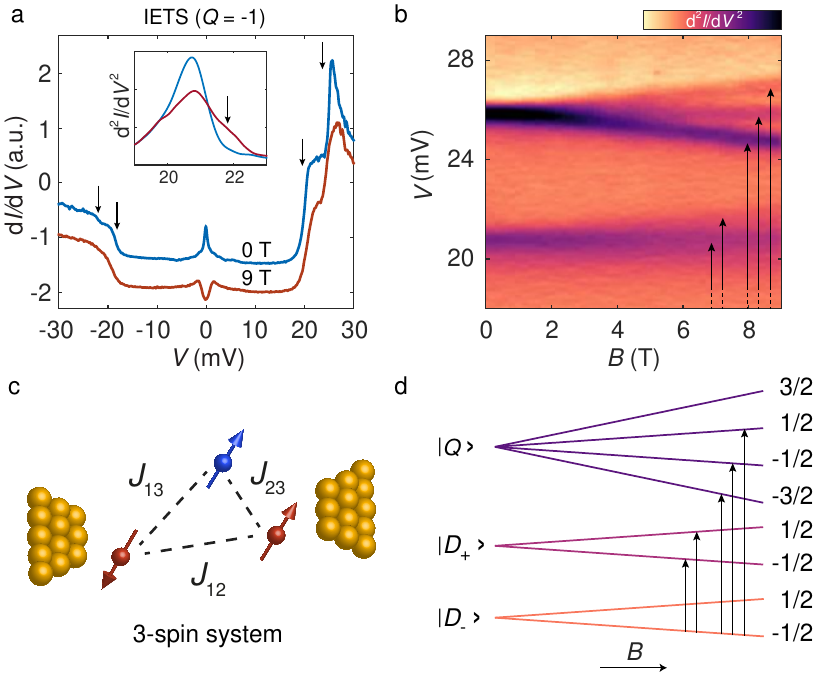}
	\caption{\textbf{Magnetic spectrum of the \refone{reduced} diradical.} (a) Differential conductance (d$I$/d$V$) spectra of the diradical junction at different magnetic fields and fixed $V_\textrm{g} = +3$ V. Two excitation steps with energies +20 mV and +25 mV split in two and three, respectively, under applied magnetic field. Inset: d$^{2}I$/d$V^{2}$ linecuts taken from (b) showing the splitting of the doublet $\left|D_{+}\right>$. (b) d$^{2}I$/d$V^{2}$ color map as a function of $V$ and $B$. (c) Schematics of a three-spin system coupled \textit{via} exchange interactions. The added electron, highlighted in blue, introduces two new exchange couplings, $J_{13}$ and $J_{23}$, with the intrinsic radical spins. (d) Spin spectra and allowed transitions for a three-spin system with antiferromagnetic $J_{12}$.}
	\label{figure4}
\end{figure}

A similar measurement is conducted at fixed $V_\textrm{g} = 3$~V. 
\Figure{figure4}(a) shows the resulting d$I$/d$V$ spectra for two different magnetic fields. At $B = 0$ T, excitation steps appear at $V = -22$ mV, $-19$ mV, $+20$ mV and $+25$ mV, together with a zero-bias peak \refone{ascribable to the Kondo effect}.\cite{Kondo1964, Liang2002} The asymmetry in bias-voltage positions with respect to $V=0$ and the different step heights can be respectively explained by a bias-dependent tuning of the exchange coupling and contributions from resonant transport with asymmetrically coupled electrodes. For increasing $B$, the zero-bias peak evolves into a dip and the excitation steps split into two (\Figure{figure4}(a), inset) and three for the low and high energy value, respectively. The d$^{2}I$/d$V^{2}$ colour map of \Figure{figure4}(b) shows this magnetic field evolution. 
From this set of excitations we deduce that the magnetic spectrum consists of a doublet ground state multiplet $\left|D_{-}\right>$ -- \refone{giving rise to the observed Kondo peak --}, a doublet excited multiplet $\left|D_{+}\right>$ and a quartet excited multiplet $\left|Q\right>$, as shown in \Figure{figure4}(d) (see \SI~ Section 1 for more details). The excitations at +20 mV and +25 mV correspond therefore to the transitions $\left|D_{-}\right>\rightarrow\left|D_{+}\right>$ and $\left|D_{-}\right>\rightarrow\left|Q\right>$, respectively. 

The spectrum we obtain at this gate voltage can only be hosted by a system like the one depicted in \Figure{figure4}(c), where the electrostatically-added electron occupies an empty orbital rather than either of the half-filled radical orbitals and couples to the two unpaired spins \textit{via} the exchange interactions $J_{13}$ and $J_{23}$.
\refone{This type of charging, observed also in two other molecular junctions of the 13 measured (see Methods for details on statistics), is in contrast to previously reported experiments on PTM monoradicals \cite{Simao2011} and other neutral diradical molecules \cite{Souto2016}.  
One of the possible explanations, explored by DFT calculations (see \SI~ Section 2.3), is that the structural distortions determining the preference for the singlet ground state lead also to a concomitant reduction of the HOMO-LUMO gap.} 

\refone{Differently than in the neutral state, the excitation spectrum of the reduced state does not provide a unique solution for $J_{12}$, $J_{13}$ and $J_{23}$, but rather a subset of solutions in the space of the three exchange couplings (see \SI~ Section 1). One scenario, obtained assuming that the coupling between the radical centers remains unchanged upon charging, yields for $J_{13}$ and $J_{23}$ the values 2 meV and 23 meV.  
In this scenario, the asymmetry between $J_{13}$ and $J_{23}$ suggests that the added spin resides in the proximity of one of the radical centers. Gas-phase DFT calculations indicate that the added electron may be delocalised over the central phenyl ring (see Section 2.4 in the \SI~).}

\begin{figure}
	\includegraphics{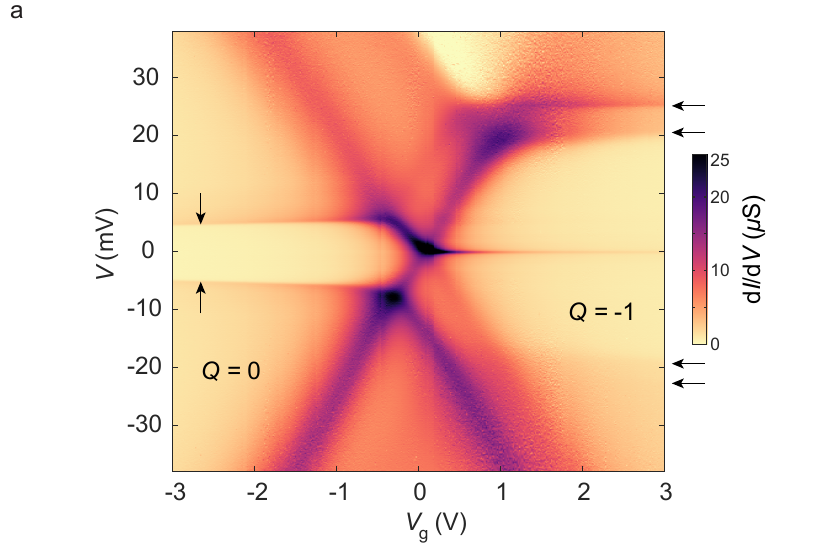}
	\caption{\textbf{The exchange-coupling gating mechanism.} (a) Differential conductance (d$I$/d$V$) as a function of $V$ and $V_\textrm{g}$ at $B = 0$ T. Slanted high-conductance edges indicate resonant transport and separate two distinct, low-conductance regions where the charge state of the molecule is stable: a neutral charge state ($Q = 0$, left) and a reduced charge state ($Q = -1$, right). Excitations lines at $\pm4.65$ mV (vertical arrows) are present in the $Q = 0$ state. In the $Q =-1$ state excitation lines appear at -19 mV, -22 mV, +20 mV and +25 mV (horizontal arrows), along with a zero-bias line of enhanced d$I$/d$V$.} 
	\label{figure3}
\end{figure}

The transport characteristics of \Figure{figure2} and \Figure{figure4} are connected as can be seen when varying the gate voltage $V_\textrm{g}$ in a continuous way. 
\Figure{figure3}(a) shows a d$I$/d$V$ map as a function of $V$ and $V_\textrm{g}$. 
The high-conductance slanted edges crossing into a zero-bias peak at $V_\textrm{g} \approx 0$~V define a resonant electron transport region separating two areas of low conductance. These features signal the presence of a single molecule in the junction whose stable charge states, labeled by $Q = 0$ and $Q = -1$, differ by one electron. The two lines in $Q = 0$, marked by vertical arrows, correspond to the singlet-to-triplet excitation steps of \Figure{figure2}, while the arrows on the right-hand side indicate the excitations discussed in \Figure{figure4}. 

The gate electrode provides thus a path to reversibly switch between the neutral and the \refone{reduced} state of the diradical molecule. Along a horizontal path around zero bias, the high-conductance peak of width $\Gamma \approx 5 $~meV is traversed. In the proximity of the peak, the molecule is in a fully mixed-valence state -- electrons from the electrodes are hopping on and off the redox center on a timescale $\tau = \hbar/\Gamma \approx 0.1$~ps. Upon application of a gate voltage $V_\textrm{g} = + 3$~V ($V_\textrm{g} = - 3$~V), within a time $\tau$ the redox center acquires a discrete occupation number and a stable spin $s = 1/2$ ($s = 0$). The presence of the spin on the redox center turns on two of the three magnetic couplings, $J_{13}$ and $J_{23}$, which, in turn, influence the time evolution of the two-spin system. This fast, electrically-controlled switching of the intramolecular magnetic interactions constitutes the essential ingredient of the quantum $\sqrt{\textrm{SWAP}}$ gate detailed in Refs. \cite{Lehmann2007, Ferrando2016a, Ferrando2016b} where two alternative read-out mechanisms are also proposed.

\section{CONCLUSIONS}

In summary, we show that incorporating an organic \refone{neutral} diradical molecule in a three-terminal device allows for reversible and stable charging from the neutral state to its \refone{reduced} state by means of a gate voltage. By performing IETS on both redox states, we find that the electron added onto the redox center magnetically couples to the radical spins, thereby driving the two-spin singlet into a three-spin doublet ground state (with three exchange couplings). In this way, by controlling the occupation of the redox center, the exchange interactions between the two radical spins and the added electron are switched on and off. Due to the large coupling to the leads, this switching takes place within sub-picosecond timescales.  

\section{METHODS}

\subsection{Details on the molecule}

The studied molecule is a neutral 2,4,6-trichloro-$\alpha,\alpha,\alpha',\alpha',\alpha'',\alpha''$-hexakis(pentachlorophenyl)mesitylene diradical prepared as previously reported\cite{Veciana1993}. Electron Spin Resonance spectroscopy in frozen solutions containing the molecules show a $S$ = 1 high-spin ground state, indicative of ferromagnetic exchange interactions between the two radical carbons in the molecule.

\subsection{Junction preparation}

The molecular solution is prepared in a water-free glove-box environment. A small amount of molecular powder is dissolved in nitrogen-saturated dichlorobenzene to a concentration of 0.5 mM approximately.

The molecular solution is deposited by drop-casting onto a Si/SiO$_2$ chip containing several Au bridges 100 nm wide, 400 nm long and 12 nm thick on top of an AuPd/Al$_2$O$_3$ gate. The nanometer-spaced source-drain electrodes are produced by feedback-controlled electromigration of these bridges\cite{Burzuri2015}. The electromigration process is stopped when the bridge conductance reaches 3-4 $G_0$. The wire is thereafter let self-break at room temperature.

A total of 160 junctions were measured, 13 of which showed signatures characteristics of spin-dependent molecular transport. Eleven of these 13 showed clear singlet-triplet excitations with antiferromagnetic coupling ranging from 0.1 meV to about 11 meV; one showed triplet-singlet characteristic with a ferromagnetic coupling of 2 meV. Four out of the 11 exhibited a degeneracy point and thus charging within the available gate voltage window. In 3 of the 4, the added charge modulates the magnetic properties. 

\subsection{Experimental conditions}

All the measurements reported in the manuscript are performed in a high-vacuum chamber ($P<5\cdot10^4$ mbar) of a dilution refrigerator ($\approx$70 mK). A built-in superconducting magnet can be used to apply magnetic fields up to 9 T.

Electrical current $I$ measurements are performed applying a DC bias voltage $V$ to the source and drain gold electrodes and/or a DC gate voltage $V_g$ while recording $I$. The differential conductance d$I$/d$V$ is obtained by taking the numerical derivative of $I$.

\begin{acknowledgement}

We acknowledge financial support by the Dutch Organization for Fundamental research (NWO/FOM), an advanced ERC grant (Mols@Mols) and the Netherlands Organisation for Scientific Research (NWO/OCW) as part of the Frontiers of Nanoscience program. EB thanks funds from the EU FP7 program through Project 618082 ACMOL, and a NWO-VENI fellowship. CR and JV thank funds from Networking Research Center on Bioengineering, Biomaterials, and Nanomedicine (CIBER-BBN), MINECO, Spain (CTQ2013-40480-R, CTQ 2016-80030-R, and "Severo Ochoa" Programme for Centers of Excellence in R$\&$D, SEV-2015-0496), MCSA ITN Network i-Switch (GA 642196), and Generalitat de Catalunya (2014-SGR-17).

\end{acknowledgement}

\clearpage

{\noindent\LARGE{\textbf{ Supporting Information for Redox-Induced Gating of the Exchange Interactions in a Single Organic Diradical}} }
\vspace{0.6cm} \\

This supplemental material is divided in two sections: Section 1 contains additional details about the fits to the magnetic spectra. The Density Functional Theory (DFT) calculations on PTM diradical molecules are presented in Section 2.

\section{1.- Fits of the magnetic excitation spectra}
\label{sec:fits}
\subsection{1.1.- Spin exchange coupling in a 2-spin system: the neutral diradical.}

In Fig. 2a in the main manuscript we show fits of the d$I$/d$V$ excitation spectra in the 2-spin state of the PTM diradical molecule. These fits are obtained using the tunneling model introduced in \cite{Ternes2015} . Within the framework of this model we describe the system by two magnetic centers with $S=1/2$ interacting through a Heisenberg exchange coupling $J_{12}$. Depending on the sign of $J_{12}$ this 2-spin system can host two distinct ground states: a singlet $\ket{S}$ and a triplet $\ket{T}$. From the excitation spectra we find the triplet to be the excited state, which implies that the exchange coupling is antiferromagnetic (positive $J_{12}$). The excitation energy for the transition $\ket{S} \rightarrow \ket{T}$ equals the exchange coupling $J_{12}$, which we determine by the fit to be 4.65 meV. In order to account for the broadening of the excitations steps, we take an effective temperature of 1.4 K. The splitting of the steps with increasing magnetic field is well reproduced by a Zeeman interaction with a g-factor of 2. A small added linear slope corrects for a possible non-flat density of states in the electrodes. 

The peaks on top of the excitation steps around $\pm 5$ mV can be reproduced in the model by two distinct mechanisms. The first being third-order tunneling processes, which yield peaks at bias voltages corresponding to the energy of the intermediate state in the scattering process (as described in ref. \cite{Ternes2015}). Alternatively, the peaks can be reproduced by including non-equilibrium effects, which follow from non-zero occupations of the excited states \cite{Ternes2015}. Upon including these effects, the height of the excitation peaks decreases with increasing energy, which is in accordance with the observed decreasing height of the split excitation steps towards higher bias. In contrast, third-order tunneling processes yield equal peak heights for the three split excitation steps. We conclude that the best agreement with the data is found by including scattering processes of second order only.

\subsection{1.2.- Spin-exchange couplings in a 3-spin system: the reduced diradical}

The excitation spectrum of the 3-spin state in the reduced diradical (Fig. 3 in the main manuscript) shows additional features which cannot be captured within the framework of the employed tunneling model. A zero-bias peak and a sharply peaked excitation step at 25 mV signal the presence of Kondo correlations, for which more involved theoretical treatments are necessary\cite{Paaske2006}. In addition, the strong bias asymmetry for energies above the excitation energies cannot be explained by second-order co-tunneling processes. The bias asymmetry is a characteristic feature of sequential electron tunneling (SET) with asymmetrically coupled electrodes. Taking these processes into account is beyond the scope of this analysis. Still, we can extract valuable information from the spectrum within the framework of a simpler model in order to make reasonable estimates of the three spin-exchange couplings of this charge state.

For the analysis of the spin excitation spectra we model the 3-spin system by the phenomenological Heisenberg-Dirac-Van Vleck (HDVV) Hamiltonian
\begin{equation}
\hat{\mathcal{H}}^{\text{HDVV}} = J_{12} \hat{\mathbf{S}}_{1} \cdot \hat{\mathbf{S}}_{2} + J_{13} \hat{\mathbf{S}}_{1} \cdot \hat{\mathbf{S}}_{3} + J_{23} \hat{\mathbf{S}}_{2} \cdot \hat{\mathbf{S}}_{3}
\end{equation}
where $J_{ij}$ represents the spin-exchange coupling between spins $i$ and $j$, and $\hat{\mathbf{S}}_{i}$ the spin operator of spin $i$. This system can host three different spin multiplets: one quartet ($\ket{Q}$) and two doublets ($\ket{D_{+}}$ and $\ket{D_{-}}$), which can be written as
\begin{equation}
\label{eq:vecquartet}
\ket{Q} = \left\{ \begin{matrix} \ket{\uparrow \uparrow \uparrow} & m = \frac{3}{2} \\ ( \ket{\uparrow \uparrow \downarrow} + \ket{\uparrow \downarrow \uparrow} + \ket{\downarrow \uparrow \uparrow})/\sqrt{3} & m = \frac{1}{2} \\ ( \ket{\uparrow \downarrow \downarrow} + \ket{\downarrow \uparrow \downarrow} + \ket{\downarrow \downarrow \uparrow})/\sqrt{3} & m = -\frac{1}{2} \\ \ket{\downarrow \downarrow \downarrow} & m = -\frac{3}{2} \end{matrix}\right. 
\end{equation}
and 
\begin{equation}
\label{eq:vecdoublet}
\ket{D_{\pm}} = \left\{ \begin{matrix} (\alpha_{123}^{\pm} \ket{\uparrow \uparrow \downarrow} + \alpha_{132}^{\pm} \ket{\uparrow \downarrow \uparrow} + \ket{\downarrow \uparrow \uparrow})/\sqrt{(\alpha_{123}^{\pm})^2+(\alpha_{132}^{\pm})^2 + 1} & m = \frac{1}{2} \\ (\alpha_{321}^{\pm} \ket{\uparrow \downarrow \downarrow} + \alpha_{312}^{\pm} \ket{\downarrow \uparrow \downarrow} + \ket{\downarrow \downarrow \uparrow})/\sqrt{(\alpha_{321}^{\pm})^2+(\alpha_{312}^{\pm})^2 + 1} & m = -\frac{1}{2} \end{matrix}\right.
\end{equation}
where the coefficients $\alpha_{ijk}^{\pm}$ depend on the values of the exchange couplings:
\begin{equation}
\alpha_{ijk}^{\pm} = \frac{J_{ij}-J_{jk} \pm X}{J_{ik}-J_{ij}}
\end{equation}
with
\begin{equation}
\label{eq:X}
X= \sqrt{J_{12}^2 + J_{13}^2 + J_{23}^2 - J_{12}J_{13} - J_{12}J_{23} - J_{13}J_{23}}.
\end{equation}
These equations show that the coefficients of the doublet eigenstates are functions of the three exchange couplings. This is in contrast with the quartet eigenstate of the three-spin system (equation \ref{eq:vecquartet}) and the singlet and triplet eigenstates of the two-spin system, which only involve numerical coefficients. The $J$-dependence disappears once a symmetry is imposed, like $J_{12}=J_{23}$ or any permutation of this equality. Here, we treat the most general case, which remains valid for $J_{12} \neq J_{13} \neq J_{23}$. The eigenenergies that correspond to the spin eigenstates are given by
\begin{equation}
\label{eq:quartet}
E_{Q} = (J_{12} + J_{13} + J_{23})/4
\end{equation}
and
\begin{equation}
\label{eq:doublet}
E_{D_{\pm}} = -(J_{12} + J_{13} + J_{23})/4 \pm X/2.
\end{equation}

\begin{figure}[t]
	\centering
	\includegraphics[width=0.7\textwidth]{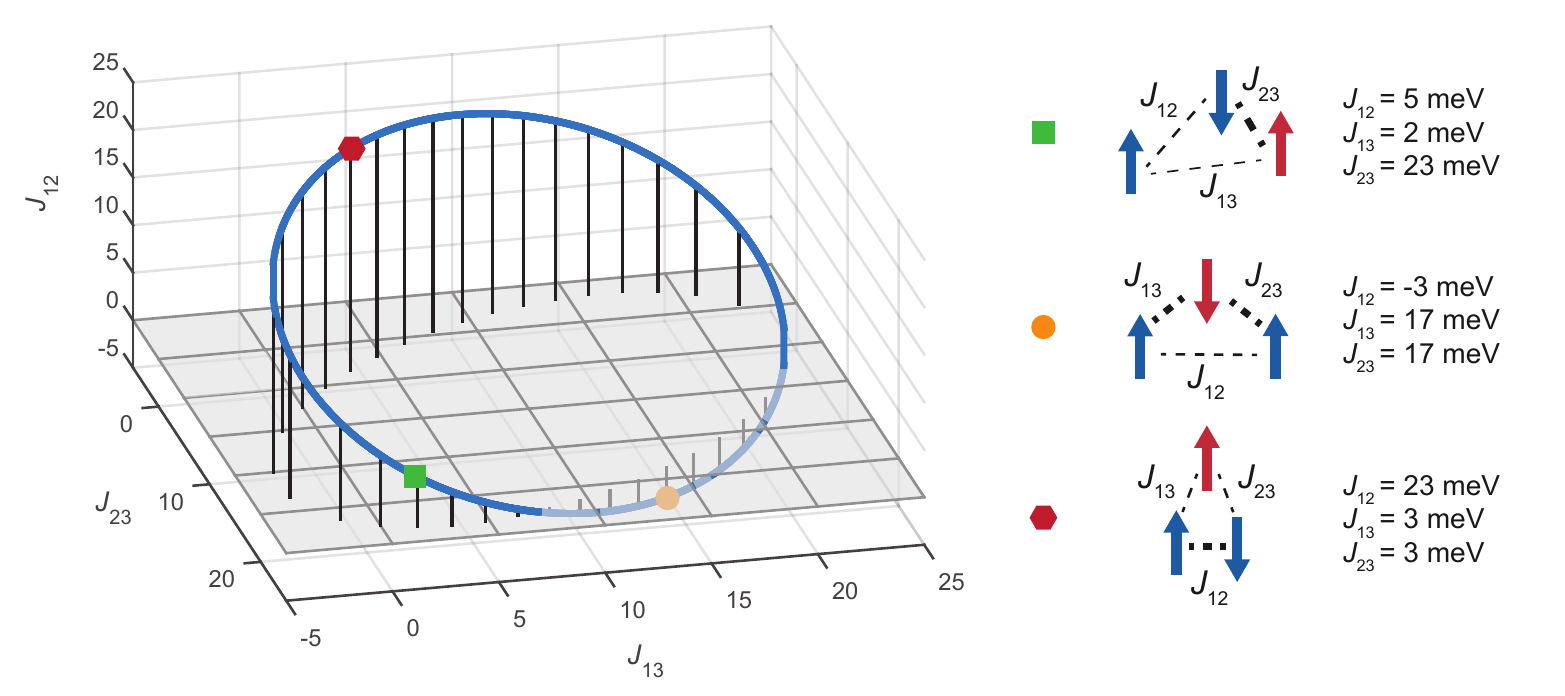}
	\caption{Solution set (blue ellipse) of equations \ref{eq:deltaE1} and \ref{eq:deltaE2} in the parameter space spanned by the exchange couplings $J_{12}$, $J_{13}$ and $J_{23}$ for the experimental values $\Delta_{1}=20$ meV and $\Delta_{2}=25$ meV. Three characteristic scenarios are highlighted: (1, green square) Coupling $J_{12}$ remains unchanged upon charging and the added spin 3 couples asymmetrically to spins 1 and 2. (2, orange circle) One coupling is weakly ferromagnetic and the other two couplings are strongly antiferromagnetic and of similar strength. (3, red hexagon) Coupling $J_{12}$ increases significantly upon charging and the added spin 3 is weakly coupled to both spin 1 and 2.}
	\label{solutions}
\end{figure}

We note that $X\geq0$, from which it follows that $\ket{D_{-}}$ is always the doublet with the lowest energy. Given that the quartet appears as an excited state in the spin spectroscopy measurements (see Fig. 3 in the main manuscript), we conclude that $\ket{D_{-}}$ is the ground state of the 3-spin system and the observed multiplet excitations correspond to $\ket{D_{-}} \rightarrow \ket{D_{+}}$ and $\ket{D_{-}} \rightarrow \ket{Q}$. The excitation energies of these transitions follow from equations \ref{eq:quartet} and \ref{eq:doublet}:  
\begin{equation}
\label{eq:deltaE1}
\Delta_{1} \equiv E_{D_{+}} - E_{D_{-}} = X, 
\end{equation}
\begin{equation}
\label{eq:deltaE2}
\Delta_{2} \equiv E_{Q} - E_{D_{-}} = (J_{12} + J_{13} + J_{23})/2 + X/2. 
\end{equation}
The experimental values we find for $\Delta_{1}$ and $\Delta_{2}$ are 20 meV and 25 meV, respectively (see Fig. 3 in the main manuscript). By equating these values with equations \ref{eq:deltaE1} and \ref{eq:deltaE2} we obtain a system of two equations with three unknown variables ($J_{12}$, $J_{13}$ and $J_{23}$). This system of equations is undetermined and its solutions lie on an ellipse in the parameter space spanned by the three exchange couplings as represented in \Fig{solutions}). Without additional knowledge about the system we cannot discern between these different solutions. 

We can, however, conclude that no symmetric solution, i.e., $J_{12} \approx J_{13} \approx J_{23}$, is available. This suggests that the added electron in the reduced charge state does not go to the position of the third radical center in the structurally equivalent PTM triradical. Moreover, if we assume that the exchange coupling of the neutral charge state remains unchanged upon charging ($J_{12} \approx 5$ meV), the solution set reduces to a single solution in which the added electron is asymmetrically coupled to the two radical centers ($J_{13} \approx 2$ meV and $J_{23} \approx 23$ meV, green square in \Fig{solutions}) and is therefore likely to be located on a ligand attached to one of the two radical centers. Two other characteristic scenarios within the solution set (red hexagon and orange circle) are highlighted in \Fig{solutions}.

Additional information can be gained from the d$I$/d$V$ spectra by analyzing the relative step height of the two spin multiplet excitations. The intensity (step height) of a transition $\ket{\psi_{i}} \rightarrow \ket{\psi_{f}}$ in the d$I$/d$V$ spectrum is proportional to the modulus squared of its transition matrix element \cite{Ternes2015}: 
\begin{equation}
\label{eq:matrixelements}
|M_{if}|^2 = \frac{1}{2}|\bra{\psi_{f}}\hat{S}_{-}^{(j)}\ket{\psi_{i}}|^2 + \frac{1}{2}|\bra{\psi_{f}}\hat{S}_{+}^{(j)}\ket{\psi_{i}}|^2 + |\bra{\psi_{f}}\hat{S}_{z}^{(j)}\ket{\psi_{i}}|^2.
\end{equation}
where $S_{-}^{(j)}$, $S_{+}^{(j)}$ and $S_{z}^{(j)}$ are the spin operators of spin $j$. In this expression, only spin-dependent second-order tunneling (co-tunneling) through spin $j$ is taken into account. Given that the coefficients of the doublet eigenstates depend on the values of the three exchange couplings (see equations \ref{eq:vecdoublet} to \ref{eq:X}) we can express the matrix elements of the transitions $\ket{D_{-}} \rightarrow \ket{D_{+}}$ and $\ket{D_{-}} \rightarrow \ket{Q}$ in terms of $J_{12}$, $J_{13}$ and $J_{23}$. Accordingly, we calculate the expected relative step height for the combinations of exchange couplings that give the observed excitation energies (the solutions shown in \Fig{solutions}). \Tab{tab:ratio} lists the relative step height for three combinations of exchange couplings.

\begin{table}[h!]
	\centering
	\caption{Ratio of step heights $D_{-} \rightarrow D_{+}$ and $D_{-} \rightarrow Q$ excitations for different combinations of exchange couplings.}
	\label{tab:ratio}
	\begin{tabular}{ccc|c}
		$J_{12}$ & $J_{13}$ & $J_{23}$ & $|M_{D_{-}\rightarrow{D_{+}}}|^2$/$|M_{D_{-}\rightarrow{Q}}|^2$  \\ \cline{1-4}
		5        & 2        & 23       & 2      \\
		-3       & 17       & 17       & 1.5    \\
		23       & 3        & 3        & 0.5    \\
	\end{tabular}
\end{table}

For the estimation of the step heights from the experimental spectrum we focus on the negative bias voltage side, where no contributions from Kondo correlations are visible and the flat excitation steps are indicative of second-order tunneling processes only. We estimate the ratio of the step heights of the two excitations to be $\sim$2, for which the present analysis favors the first configuration in Table \ref{tab:ratio}. Two spins are strongly coupled to each other ($\sim$23 meV), whereas the third spin is relatively weakly coupled (2-5 meV) to the former two. This is consistent with the scenario we proposed before, in which $J_{12} \approx 5$ meV as in the 2-spin state, $J_{13} \approx 2$ meV and $J_{23} \approx 23$ meV.

The contribution of elastic co-tunneling however, which we have neglected so far, is in this configuration one order of magnitude larger than the inelastic co-tunneling steps, which is not what we observe in the experimental spectrum. In fact, if we consider transport through only one spin there is no combination of exchange couplings that reproduces the measurements. Only by introducing an interfering channel through a second spin, both the ratio of the inelastic step heights, as well as the observed elastic co-tunneling contribution match the experimental spectrum. The transport through this three-spin system occurs therefore through at least two interfering channels.

\section{2.- Details on the DFT calculations}
\label{sec:DFT}
\subsection{2.1.- Methodology employed}
\subsubsection{2.1.1.- Formalism, basis sets and program used}

We have additionally performed calculations on the PTM diradical system within the spin unrestricted Density Functional Theory (DFT) formalism, using the well-known B3LYP\cite{Becke1993} hybrid functional and the Pople-type basis set 6-31G(d,p)\cite{Hehre1972,Dill1975,Francl1982}for all atoms, as implemented in the Gaussian09$\_d$\cite{Frisch2009} suite of programs.

\subsubsection{2.1.2.- Calculation of magnetic coupling constants in the diradical molecule: Yamaguchi's formula}

Throughout this work, we have assumed a model spin Hamiltonian defined as $\textit{\^{H}}^{HDVV}=\Sigma_{\langle i,j \rangle}J_{ij}\textbf{S}_{i}\cdot\textbf{S}_{j}$, where a negative $J$ value indicates a ferromagnetic interaction of the unpaired electrons, whereas a positive $J$ denotes an antiferromagnetic interaction. This corresponds with a situation in which both unpaired electrons are aligned parallel and antiparallel, respectively. The calculation of the magnetic coupling constants in the neutral diradical molecule have been done using the formula proposed by Yamaguchi\cite{Yamaguchi1987,Yamaguchi1988,Yamaguchi1988a}, where the triplet (T) spin adapted state is approximated by a high-spin Kohn-Sham determinant with two unpaired electrons with parallel spins and the singlet (S) spin adapted state by a broken symmetry (BS) solution\cite{Noodleman1981,Noodleman1986,Noodleman1995}. In this way, the vertical DFT triplet-singlet gap is approximately twice the energy difference between the high spin and the BS solutions, 

\begin{equation}
\Delta_{vert}= E_S-E_T=\frac{2\left(E_{BS}-E_{T}\right)}{\langle S^2_T \rangle - \langle S^2_{BS}\rangle}
\label{formula1}
\end{equation}

\noindent where $E_s$, $E_T$ and $E_{BS}$ are the energies of the singlet, triplet and broken symmetry states, respectively. The denominator contains the expectation value of the square of the total spin operator for the triplet and BS solutions (close to 2.000 and 1.000, respectively). 

\subsection{2.2.- Exploration of distortions leading to ground state multiplicity reversal}

For the most extensively investigated sample, the ground state is a singlet state (see Fig. 2 in the main manuscript), as opposed to the triplet ground state of the same molecule in frozen solution\cite{Veciana1993}. Thus, it is reasonable to think that the reversal in the ground state multiplicity of the molecule comes from its interaction with the electrodes\cite{Gaudenzi2016}. Following a very similar reasoning to the one presented for the related PTM triradical case\cite{Gaudenzi2016}, we have studied some sensible potential energy surfaces (PES) in order to check whether a torsion of the main dihedral angles also lead to a change in the sign of the magnetic exchange coupling of the molecule.

\subsubsection{2.2.1.- Definition of the structural parameters modified and associated geometries. Distortion along D6 and D24 dihedral angles.}

The z-matrix used to investigate the effect of the distortion on the electronic structure of the molecule is presented in Table \ref{tab:tableSD1}.

\begin{longtable}[h!]{cc|cc|cc|cc}
	\caption{z-matrix corresponding to the PTM diradical}\\
	\hline \\
	\label{tab:tableSD1}
	
	C     \\         
	C        & 1   &    B1 \\
	C        & 2   &    B2  &  1   &      A1 \\
	C        & 3   &    B3  &  2   &      A2  &  1   &    D1  &  0 \\  
	C        & 4   &    B4  &  3   &      A3  &  2   &    D2  &  0 \\
	C        & 5   &    B5  &  4   &      A4  &  3   &    D3  &  0 \\
	C        & 6   &    B6  &  5   &      A5 &   4   &    D4  &  0 \\
	C        & 4   &    B7  &  3   &      A6 &   2   &    D5 &   0 \\
	C        & 7   &    B8  &  6   &      A7 &   5   &    D6  &  0 \\
	C        & 9   &    B9  &  7   &      A8 &   6   &    D7  &  0 \\
	C        & 9   &    B10 &  7   &      A9  &  6   &         D8  &  0 \\
	C        & 10  &    B11 &  9   &      A10  &  7   &         D9  &  0 \\
	C        & 11  &    B12 &  9   &      A11  &  7   &        D10  &  0 \\
	C        & 13  &    B13 &  11  &      A12 &   9    &       D11  &  0 \\
	C        & 7   &    B14 &  6   &      A13  &  5    &       D12  &  0 \\
	C        & 15  &    B15 &  7   &      A14  &  6     &      D13  &  0 \\
	C        & 15  &    B16 &  7   &      A15 &   6     &      D14 &   0 \\
	C        & 16  &    B17 &  15  &      A16 &   7    &       D15  &  0 \\
	C        & 17  &    B18 &  15  &      A17 &   7     &      D16  &  0 \\
	C        & 19  &    B19 &  17  &      A18 &  15    &      D17  &  0 \\
	C        & 8   &    B20 &  4   &      A19  &  3      &     D18  &  0 \\
	C        & 21  &    B21 &  8   &      A20  &  4     &      D19  &  0 \\
	C        & 21  &    B22 &  8   &      A21  &  4     &      D20  &  0 \\
	C        & 22  &    B23 &  21  &      A22 &   8    &       D21  &  0 \\
	C        & 23  &    B24 &  21  &         A23  &  8     &      D22  &  0 \\
	C        & 24  &    B25 &  22  &          A24 &  21    &       D23 &   0 \\
	C        & 8   &    B26 &   4  &         A25 &   3     &      D24  &  0 \\
	C        & 27  &    B27 &   8  &         A26 &   4    &       D25  &  0 \\
	C        & 27  &    B28 &   8  &         A27 &   4    &       D26 &   0 \\
	C        & 28  &    B29 &  27  &         A28 &   8    &       D27  &  0 \\
	C        & 29  &    B30 &  27  &         A29  &  8    &       D28  &  0 \\
	C        & 31  &    B31 &  29  &         A30 &  27    &       D29 &   0 \\
	Cl       & 5   &    B32 &   4  &        A31  &  3      &     D30  &  0 \\
	Cl       & 1   &    B33 &   2  &         A32 &   3    &       D31 &   0 \\
	Cl       & 3   &    B34 &   2  &         A33  &  1    &       D32  &  0 \\
	Cl       & 17  &    B35 &  15  &         A34 &   7    &       D33  &  0 \\
	Cl       & 16  &    B36 &  15  &         A35 &   7    &       D34  &  0 \\
	Cl       & 18  &    B37 &  16  &         A36 &  15    &       D35  &  0 \\
	Cl       & 20  &    B38 &  19  &         A37 &  17    &       D36  &  0 \\
	Cl       & 19  &    B39 &  17 &          A38 &  15    &       D37  &  0 \\
	Cl       & 10  &    B40 &   9  &         A39 &   7    &       D38  &  0 \\
	Cl       & 12  &    B41 &  10  &         A40 &   9    &       D39  &  0 \\
	Cl       & 14  &    B42 &  13  &         A41 &  11     &      D40  &  0 \\
	Cl       & 13  &    B43 &  11   &        A42 &   9     &      D41 &   0 \\
	Cl       & 11  &    B44 &   9  &         A43 &   7     &      D42 &   0 \\
	Cl       & 22  &    B45 &  21  &         A44 &    8    &      D43  &  0 \\
	Cl       & 24  &    B46 &  22  &         A45 &  21     &      D44  &  0 \\
	Cl       & 26  &    B47 &  24  &         A46 &  22     &      D45 &   0 \\
	Cl       & 25  &    B48 &  23  &         A47 &  21     &      D46  &  0 \\
	Cl       & 23  &    B49 &  21   &        A48 &   8     &      D47  &  0 \\
	Cl       & 29  &    B50 &  27  &         A49 &   8     &      D48  &  0 \\
	Cl       & 31  &    B51 &  29   &        A50 &  27     &      D49  &  0 \\
	Cl       & 32  &    B52 &  31  &         A51 &  29     &      D50  &  0 \\
	Cl       & 30  &    B53 &  28  &         A52 &  27     &      D51  &  0 \\
	Cl       & 28  &    B54 &  27   &        A53 &   8     &      D52 &   0 \\
	C        & 2   &    B55 &   1  &         A54 &   6     &      D53 &   0 \\
	H        & 56  &    B56  &  2   &        A55 &   1     &      D54  &  0 \\
	C        & 56  &    B57 &   2    &       A56  &  1     &      D55  &  0 \\
	C        & 58  &    B58 &  56    &       A57 &   2     &      D56  &  0 \\
	C        & 58  &    B59 &  56   &        A58 &   2     &      D57  &  0 \\
	C        & 59  &    B60 &  58    &       A59 &  56     &      D58  &  0 \\
	C        & 60  &    B61 &  58   &        A60 &  56     &      D59  &  0 \\
	C        & 61  &    B62 &  59    &       A61 &  58     &      D60  &  0 \\
	C        & 56  &    B63 &   2    &       A62 &   1     &      D61  &  0 \\
	C        & 64  &    B64 &  56    &       A63 &   2     &      D62  &  0 \\
	C        & 64  &    B65 &  56    &       A64 &   2     &      D63  &  0 \\
	C        & 65  &    B66 &  64   &        A65 &  56     &      D64  &  0 \\
	C        & 66  &    B67 &  64   &        A66 &  56     &      D65  &  0 \\
	C        & 68  &    B68 &  66   &        A67 &  64     &      D66  &  0 \\
	Cl       & 59  &    B69 &  58   &        A68 &  56     &      D67  &  0 \\
	Cl       & 61  &    B70 &  59   &        A69 &  58     &      D68  &  0 \\
	Cl       & 63  &    B71 &  61   &        A70 &  59     &      D69  &  0 \\
	Cl       & 62  &    B72 &  60   &        A71 &  58     &      D70  &  0 \\
	Cl       & 60  &    B73 &  58   &        A72 &  56     &      D71  &  0 \\
	Cl       & 65  &    B74 &  64   &        A73 &  56     &      D72  &  0 \\
	Cl       & 66  &    B75 &  64   &        A74 &  56     &      D73  &  0 \\
	Cl       & 68  &    B76 &  66   &        A75 &  64     &      D74  &  0 \\
	Cl       & 69  &    B77 &  68   &        A76 &  66     &      D75  &  0 \\
	Cl       & 67  &    B78 &  65   &        A77 &  64     &      D76  &  0 \\
	\\
	\hline \\
	B1      &       1.41110046 &
	B2      &       1.40623962 &
	B3      &       1.41978805 &
	B4      &       1.41343770\\
	B5      &       1.41378455 &
	B6      &       1.48223807 &
	B7      &       1.48498927 &
	B8      &       1.48361690\\
	B9      &       1.41721731&
	B10     &       1.41670027&
	B11     &       1.40178991&
	B12     &       1.40189910\\
	B13     &       1.40354201&
	B14     &       1.48368924&
	B15     &       1.41744118&
	B16    &        1.41673432\\
	B17    &        1.40192059&
	B18     &       1.40187313&
	B19     &       1.40342067&
	B20     &       1.48525082\\
	B21     &       1.41708194&
	B22     &       1.41726283&
	B23     &       1.40217844&
	B24     &       1.40189633\\
	B25     &       1.40342291&
	B26     &       1.48306291&
	B27     &       1.41785776&
	B28     &       1.41636146\\
	B29     &       1.40183323&
	B30     &       1.40153259&
	B31     &       1.40341220&
	B32    &        1.74831983\\
	B33    &        1.75561754&
	B34    &        1.74516520&
	B35     &       1.74009318&
	B36     &       1.74212938\\
	B37     &       1.73607710&
	B38     &       1.73232164&
	B39     &       1.73561129&
	B40     &       1.74188847\\
	B41     &       1.73600249&
	B42    &        1.73222137&
	B43     &       1.73577108&
	B44     &       1.74115301\\
	B45     &       1.74085784&
	B46     &       1.73592562&
	B47     &       1.73234834&
	B48     &       1.73601725\\
	B49     &       1.74222854&
	B50     &       1.74094026&
	B51     &       1.73597741&
	B52     &       1.73221535\\
	B53     &       1.73605932&
	B54     &       1.74195685&
	B55    &        1.54965667&
	B56     &       1.08968451\\
	B57     &       1.54689301&
	B58     &       1.41219813&
	B59     &       1.40887951&
	B60     &       1.40232451\\
	B61     &       1.40646844&
	B62     &       1.40122086&
	B63     &       1.54578100&
	B64      &      1.41279197\\
	B65     &       1.40825787&
	B66    &        1.40241982&
	B67     &       1.40663455&
	B68      &      1.40125347\\
	B69      &      1.74947567&
	B70     &       1.73562338&
	B71      &      1.73352162&
	B72      &      1.73644598\\
	B73     &       1.74098254&
	B74     &       1.74947769&
	B75     &       1.73996828&
	B76     &       1.73617894\\
	B77     &       1.73367965&
	B78     &       1.73568834&
	A1      &     116.15200149&
	A2      &     123.62176612\\
	A3      &     116.42678592&
	A4      &     123.62912345&
	A5      &     121.96838752&
	A6     &      122.23683276\\
	A7     &      120.18348452&
	A8      &     121.69331750&
	A9      &     121.67856645&
	A10     &     121.95451174\\
	A11      &    122.01207771&
	A12     &     119.87916212&
	A13     &     120.75548851&
	A14     &     121.72892671\\
	A15     &     121.70896091&
	A16     &     121.99637777&
	A17     &     122.04436666&
	A18    &      119.89860715\\
	A19     &     119.38617092&
	A20     &     121.63791729&
	A21     &     121.72174332&
	A22     &     121.96117044\\
	A23     &     121.96436358&
	A24     &     119.91786475&
	A25     &     121.71324089&
	A26     &     121.61284695\\
	A27     &     121.89800608&
	A28     &     122.01062097&
	A29     &     122.11887953&
	A30     &     119.89600982\\
	A31     &     117.95303377&
	A32    &      118.20814329&
	A33     &     119.05104219&
	A34     &     120.02776618\\
	A35     &     120.16314170&
	A36     &     120.55129186&
	A37     &     120.19896052&
	A38    &      120.51927767\\
	A39     &     120.14781038&
	A40     &     120.53217287&
	A41     &     120.20574392&
	A42     &     120.54829188\\
	A43     &     120.03862873&
	A44     &     120.11485682&
	A45     &     120.52027907&
	A46     &     120.20842394\\
	A47     &     120.54424116&
	A48     &     120.20302906&
	A49     &     119.97894621&
	A50     &     120.53866349\\
	A51     &     120.22459957&
	A52     &     120.53526188&
	A53     &     120.15692109&
	A54      &    117.23526484\\
	A55     &     100.40821185&
	A56     &     118.09905708&
	A57    &      117.20805526&
	A58     &     126.19194662\\
	A59     &     122.66565118&
	A60     &     121.84346099&
	A61    &      119.52318213&
	A62     &     116.43697742\\
	A63     &     117.17076063&
	A64     &     126.18522521&
	A65     &     122.53064102&
	A66     &     121.79272579\\
	A67    &      120.19024546&
	A68     &     119.90034361&
	A69     &     120.74960101&
	A70     &     120.23794434\\
	A71     &     120.47356697&
	A72     &     121.12888265&
	A73     &     120.02825037&
	A74     &     121.07499802\\
	A75    &      120.46948734&
	A76    &      120.39302407&
	A77     &     120.72713711&
	D1      &      -0.95174211\\
	D2      &       2.54533592&
	D3      &      -2.07628012&
	D4      &    -179.63885168&
	D5     &     -176.86447029\\
	D6     &       48.61854578&
	D7     &     -129.82217583&
	D8     &       50.77587796&
	D9     &     -178.94574921\\
	D10    &      179.91664749&
	D11    &       -0.97285620&
	D12    &     -131.34695591&
	D13    &     -130.13844772\\
	D14    &       50.52473707&
	D15    &     -178.83247176&
	D16    &      179.84521993&
	D17    &       -1.00111356\\
	D18     &     129.39427352&
	D19    &      -50.83335062&
	D20    &      129.88663290&
	D21    &      179.97133298\\
	D22    &      179.13611280&
	D23    &        0.96700621&
	D24    &      -50.35746961&
	D25    &      130.29156719\\
	D26    &      -50.54563294&
	D27    &      178.21636742&
	D28    &     -179.24456123&
	D29    &        0.89355281\\
	D30    &      170.68688686&
	D31    &      175.06115726&
	D32    &     -179.78679052&
	D33    &        3.68286349\\
	D34    &        5.05661448&
	D35    &     -179.92566475&
	D36    &     -179.51017760&
	D37    &     -179.88190289\\
	D38    &        4.91389451&
	D39   &      -179.86300760&
	D40   &      -179.48979924&
	D41   &      -179.80027517\\
	D42   &         3.84869270&
	D43   &        -4.17958274&
	D44   &       179.67908609&
	D45   &       179.60282890\\
	D46   &       179.75251817&
	D47   &        -4.53343892&
	D48   &        -2.74541147&
	D49   &       179.84563170\\
	D50   &       179.37868184&
	D51   &       -179.92244745&
	D52   &        -5.88960332&
	D53   &      -175.86468260\\
	D54   &        35.85665666&
	D55   &       143.36326213&
	D56   &       -72.66062300&
	D57   &       113.04533645\\
	D58   &      -174.83191476&
	D59   &       173.09710148&
	D60   &         1.15846333&
	D61   &       -71.55453258\\
	D62   &       142.70510604&
	D63   &       -33.03611950&
	D64   &      -176.18360822&
	D65   &       174.70505986\\
	D66   &         1.14056636&
	D67   &         4.92313696&
	D68   &      -179.54327640&
	D69   &       179.09651092\\
	D70   &      -179.34785660&
	D71   &        -9.91628213&
	D72   &         3.38993554&
	D73   &        -7.92182933\\
	D74   &      -179.58801802&
	D75   &       179.67427914&
	D76   &      -179.63358334&
	\\
	\hline
\end{longtable}

\noindent where the parameters D6 and D24 are used to perform the distortion. Note that the presented values correspond with the stationary point in the PES of the triplet state (the absolute minimum) and these dihedral angles define the relative position of the carbon-centered radical with the shared central ring, as indicated in \Fig{fig:SD1}.

\begin{figure}[h!]
	\includegraphics[width=0.6\textwidth]{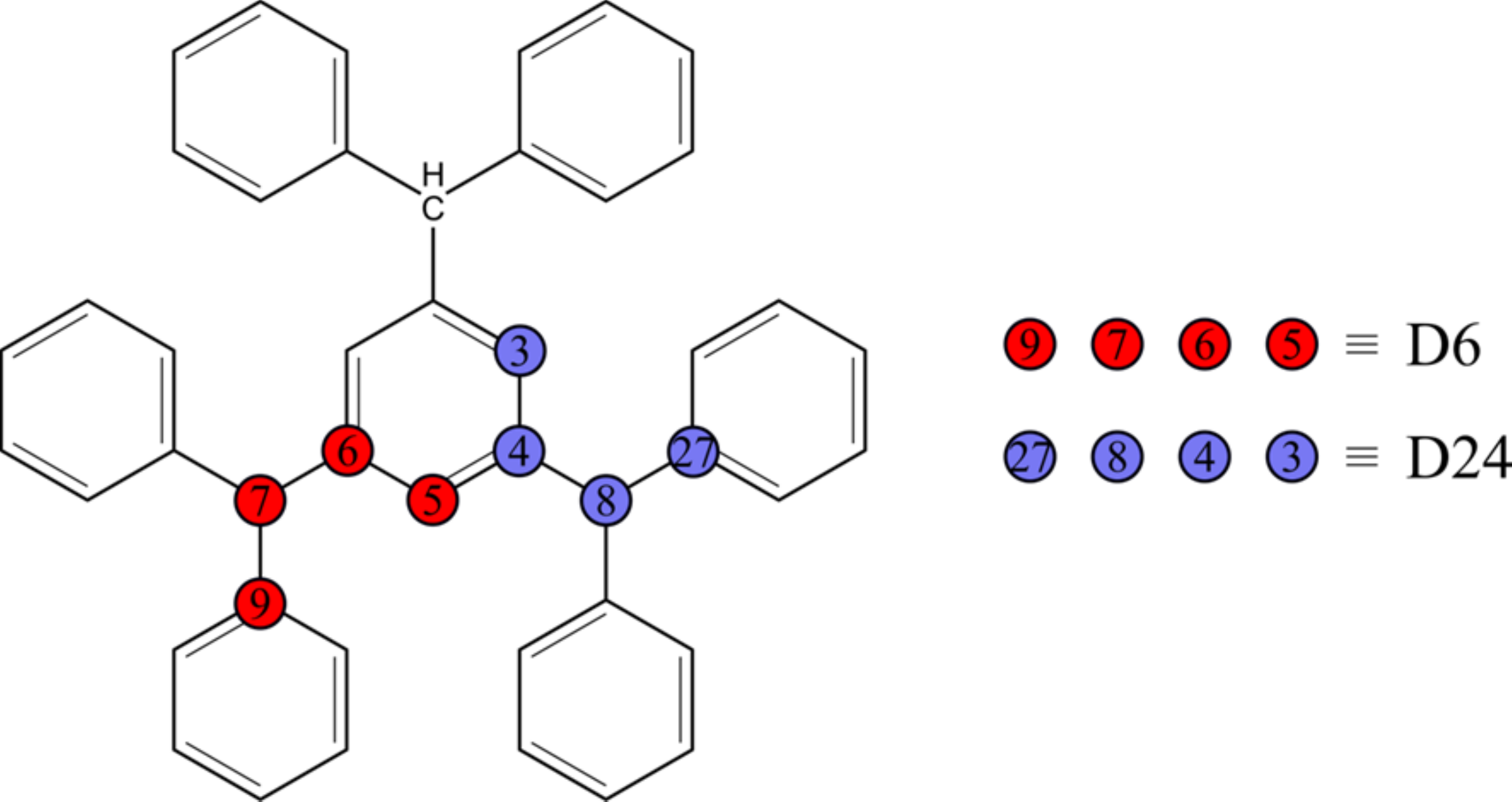}
	\caption{Graphical representation of structural parameters modified. The numbers correspond to the actual labels used in the z-matrix above. Note that the Chlorine atoms have not been represented for the sake of interpretation.}
	\label{fig:SD1}
\end{figure}

Now, in order to roughly simulate one of the possible impacts of the electrodes on these molecules, we undergo an investigation of the energies of the triplet and broken symmetry solutions at a series of distorted points. The strategy followed consists in modifying the D6 and D24 values, and allow a relaxation of the rest of structural parameters for the triplet state. Once the restricted optimization has located a stationary point (keeping D6 and D24 fixed), this geometry is used to perform a single point calculation of the broken symmetry solution. With these two states and using Yamaguchi's formula, one can calculate the magnetic coupling constant at each geometry.

\subsubsection{2.2.2.- Results: Energetic cost and magnetic coupling constant \textit{vs} distortion}

The different points in the PES of the triplet, their corresponding absolute energies, the associated energetic cost to undergo the distortion from the absolute minimum and the extracted magnetic coupling constants are displayed in Table \ref{tab:tableSD2}. 

\begin{longtable}[h!]{cccccc}
	\caption{Energy and magnetic coupling constants for different points in the PES of the triplet}\\
	\hline
	\label{tab:tableSD2}
	D6, D24                                     & State & Energy  (a.u.)         & $\langle S^2 \rangle$     & $\Delta E$ (Kcal/mol)   & $J$ (meV) \\
	\hline
	\multirow{2}{*}{35,-35}                     & T     & -16901.5716995 & 2.052  & 2.4 &         \\
	& BS    & -16901.5698462 & 1.0186 &     & -97.6    \\
	\multicolumn{1}{c}{\multirow{2}{*}{40,-40}} & T     & -16901.5738229 & 2.049  & 1.1 &         \\
	\multicolumn{1}{c}{}                        & BS    & -16901.5723603 & 1.0193 &     & -77.3    \\
	\multirow{2}{*}{45,-45}                     & T     & -16901.5751313 & 2.047  & 0.3 &         \\
	& BS    & -16901.5739896 & 1.0198 &     & -60.5    \\
	\multirow{2}{*}{48.62,-50.36}               & T     & -16901.5755320 & 2.045  & 0.0 &         \\
	& BS    & -16901.5746406 & 1.020  &     & -47.4    \\
	\multirow{2}{*}{55,-55}                     & T     & -16901.5749341 & 2.043  & 0.4 &         \\
	& BS    & -16901.5743061 & 1.0206 &     & -33.4    \\
	\multirow{2}{*}{60,-60}                     & T     & -16901.5733634 & 2.042  & 1.4 &         \\
	& BS    & -16901.5729504 & 1.0209 &     & -22.0    \\
	\multirow{2}{*}{65,-65}                     & T     & -16901.5707591 & 2.041  & 3.0 &         \\
	& BS    & -16901.5705208 & 1.0212 &     & -12.7    \\
	\multirow{2}{*}{70,-70}                     & T     & -16901.5670913 & 2.040  & 5.3 &         \\
	& BS    & -16901.5669974 & 1.0220 &     & -5.0     \\
	\multirow{2}{*}{75,-75}                     & T     & -16901.5623530 & 2.040  & 8.3 &         \\
	& BS    & -16901.5623853 & 1.0231 &     & 1.7  \\
	\hline 
\end{longtable}

\begin{figure}[h!]
	\includegraphics[width=0.6\textwidth]{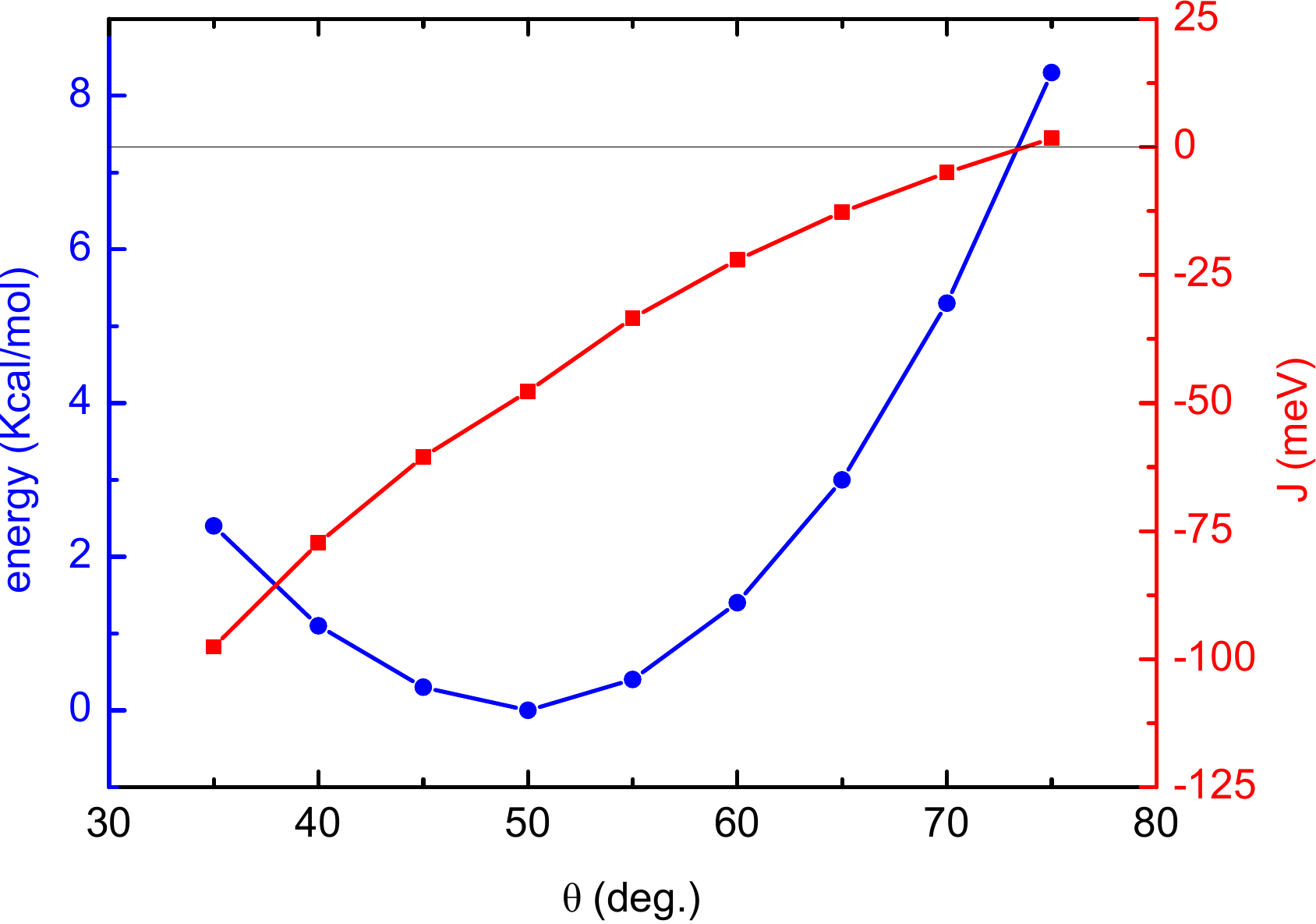}
	\caption{Graphical representation of the energetic cost and associated magnetic coupling constant at each restricted optimized geometry. Data points are extracted from Table S3.}
	\label{fig:SD2}
\end{figure}

The graphical representation of these values is presented in \Fig{fig:SD2}. We observe a similar tendency as in the previously reported PTM triradical\cite{Gaudenzi2016}: a distortion of the molecule paying relatively low energies can invert the sign of the exchange coupling from ferro to antiferro.

\subsection{2.3.- Exploration of distortions leading to a concomitant ground state multiplicity reversal and a decrease of the HOMO-LUMO gap}

The experimental results discussed in the main text also indicate that the measured sample can be reversibly charged, passing from the neutral diradical to a reduced (anionic) diradical species which maintains the quartet and two doublet magnetic states in the low-lying region of the spectrum. This charging is in contrast to the monoradical\cite{Frisenda2015} and triradical\cite{Gaudenzi2016} suggests a lower HOMO-LUMO gap in the diradical case potentially induced by the distortion of the molecule as a consequence of the interaction with the electrodes. However, for both the triplet and BS solutions in each of the points discussed in \Fig{fig:SD2}, the HOMO-LUMO gap keeps a constant value of about 2 eV.

\begin{figure}[h!]
	\includegraphics[width=0.6\textwidth]{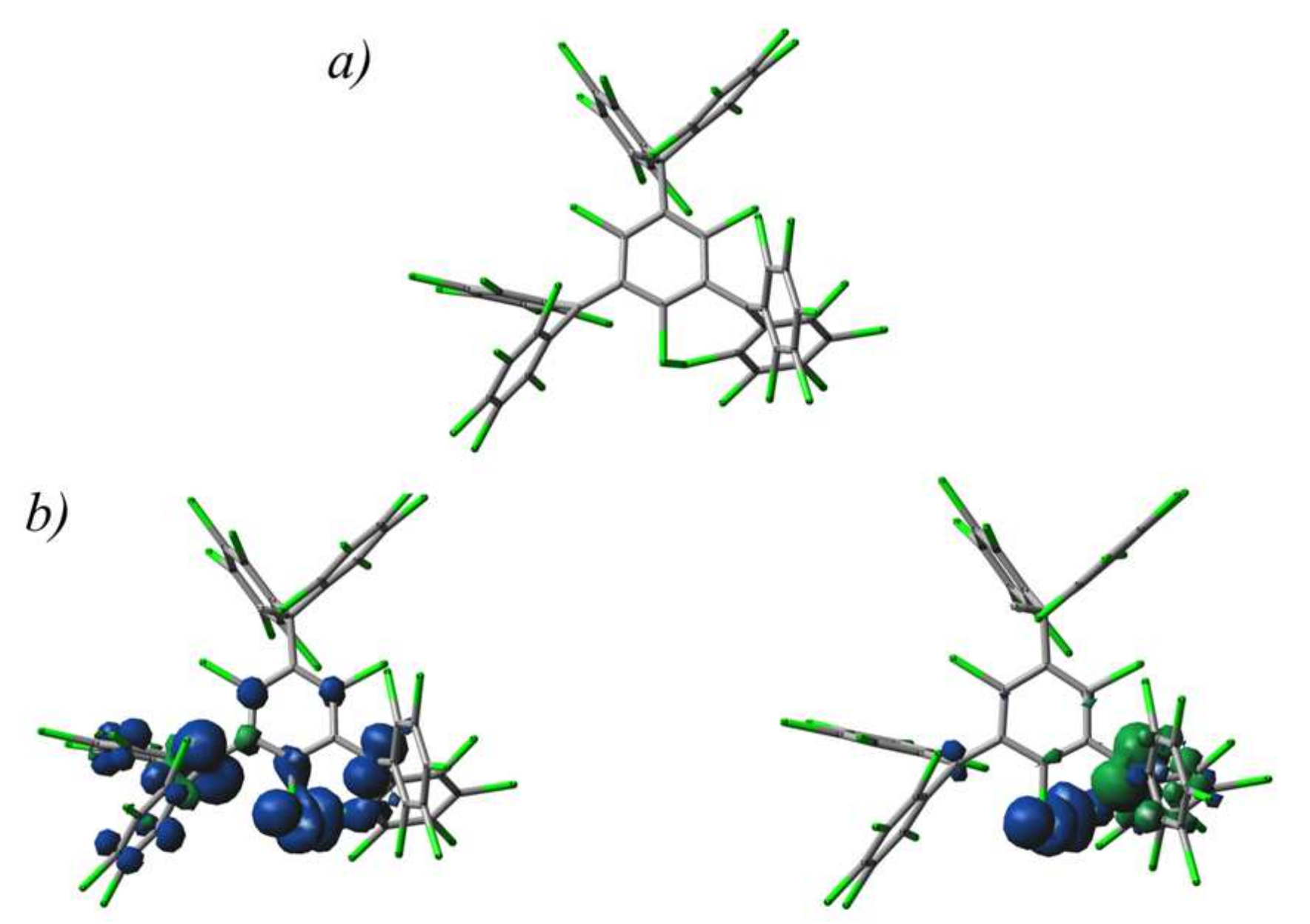}
	\caption{Representation of a) structure and b) spin densities of triplet and BS, respectively, in the situation where Cl$\cdot\cdot\cdot$Cl destroys the nature of the magnetic states}
	\label{fig:SD3}
\end{figure}

Thus, we performed a series of more extended distortions where the rest of the structural parameters were not relaxed. Again, we calculated the triplet and BS solutions for each geometry and investigated the impact on the HOMO-LUMO gap. We have found four differential behaviours (detailed in Table \ref{tab:tableSD3}): first, one in which the distortions lead to a singlet ground state but leave intact the HOMO-LUMO gap; second, a situation in which the distortions reduce the HOMO-LUMO gap but result in even larger energy differences between the ground triplet and excited BS solutions; third, corresponding to the case that would explain the experimental results, where the distortion concomitantly stabilizes the BS state as the ground state and reduces the HOMO-LUMO gap. Finally, we have also observed that if the distortions are too large, to the point of having very close Cl$\cdot\cdot\cdot$Cl or Cl$\cdot\cdot\cdot\pi$-system interactions, the nature of the magnetic states is not maintained and the spin density is no longer dominated by the carbon-based radical centres (see \Figure{fig:SD3}). In this case, the HOMO-LUMO gap of the much more  stable BS solutions is reduced to~0.5 eV. However, since the experimental data clearly resolves the singlet-triplet spectrum, the latter case is discarded as a plausible explanation. 

\subsubsection{2.3.1.- Definition of the modified structural parameters and associated geometries}

\begin{figure}[h!]
	\includegraphics[width=0.6\textwidth]{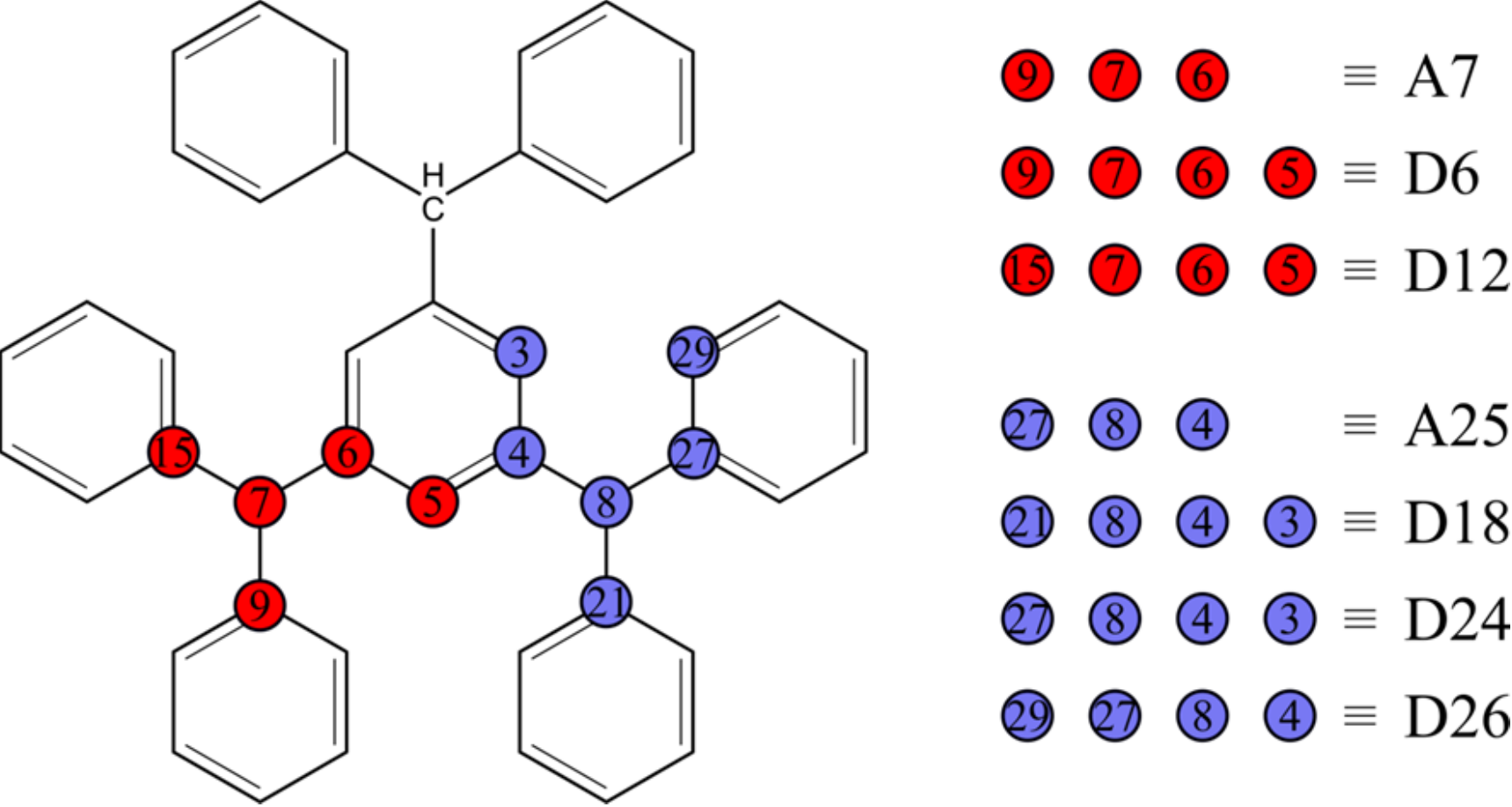}
	\caption{Graphical representation of structural parameters modified. The numbers correspond to the actual labels used in the z-matrix above. Note that the Chlorine atoms have not been represented for the shake of interpretation.}
	\label{fig:SD4}
\end{figure}

Using the same definition of the z-matrix as the one presented in Table \ref{tab:tableSD1}, we investigated several PES by modifying a series of different structural parameters. \Figure{fig:SD4} depicts the new set of parameters that have been considered.

\subsubsection{2.3.2.- Results: Magnetic coupling constant and HOMO-LUMO gap \textit{vs} distortion.}

Table \ref{tab:tableSD3} presents the different PES investigated with the corresponding values for each structural parameter modified, the energy difference between the triplet and BS solutions and the calculated HOMO-LUMO gap at each geometry. As evidenced, there is a clear relationship between distortion and HOMO-LUMO gap, although it is not simple. The lowest value for the HOMO-LUMO gap that we could get was 1.5 eV for a BS ground state. Despite not being small enough to account for the capacity of the sample to reversibly charge, these results clearly indicate that the HOMO-LUMO gap can be modified by simple structural distortions. And despite not having located the particular region in the PES of the molecule where that could happen, the presented results establish that this can indeed be achieved. 

\begin{longtable}[h!]{c|c|c|c}
	\caption{Summary of $\Delta E_{T-BS}$ (meV) and $\Delta E_{HOMO-LUMO}$ (eV) for each geometry of the different PES investigated. Note that a positive value of $\Delta E_{T-BS}$ implies a more stable BS solution. Also, the first and second entries in the $\Delta E_{HOMO-LUMO}$ column stand for the values calculated for triplet and BS, respectively. The last row indicates the values of modified parameters at the absolute minimum.}
	\label{tab:tableSD3}\\
	\multicolumn{4}{l}{Structural parameters modified}                                                 \\
	\hline
	label                                       & values                           & $\Delta E_{T-BS}$   & $\Delta E_{H-L}$   \\
	&                                  &   (meV)    &    (eV)\\          
	\hline
	\multirow{11}{*}{D6, D24}                   & 10, -50.36                       & -39.4 & 1.9 / 2.1 \\
	& 30, -50.36                       & -32.2 & 2.2 / 2.3 \\
	& 50, -50.36                       & -22.8 & 2.3 / 2.4 \\
	& 70. -50.36                       & -13.0 & 2.3 / 2.5 \\
	& 90, -50.36                       & -5.1  & 2.3 / 2.4 \\
	&                                  &       &           \\
	& 10, -90                          & -8.6  & 1.8 / 1.9 \\
	& 30, -90                          & -2.6  & 2.1 / 2.2 \\
	& 50, -90                          & +1.1  & 2.2 / 2.4 \\
	& 70. -90                          & +2.9  & 2.2 / 2.4 \\
	& 90, -90                          & +3.7  & 2.2 / 2.4 \\
	&                                  &       &           \\
	\hline
	\multirow{5}{*}{D6, D12, D18, D24}          & -180, 0, 90, -90                 & -5.1  & 1.1 / 1.1 \\
	& -150, 30, 90, -90                & -2.1  & 1.6 / 1.6 \\
	& -120, 60, 90, -90                & +14.5 & 1.7 / 1.8 \\
	& -90, 90, 90, -90                 & +43.2 & 1.9 / 2.2 \\
	& -0, -180, 90, -90                & -4.7  & 1.1 / 1.1 \\
	&                                  &       &           \\
	\hline
	\multirow{5}{*}{A7, D6, D12, D18, D24}      & 150, -48.6, -131.3, 90, -90      & +9.5  & 1.7 / 1.7 \\
	& 140, -48.6, -131.3, 90, -90      & +9.2  & 1.7 / 1.8 \\
	& 120, -48.6, -131.3, 90, -90      & +7.3  & 1.7 / 1.7 \\
	& 100, -48.6, -131.3, 90, -90      & +2.2  & 1.5 / 1.5 \\
	& 90, -48.6, -131.3, 90, -90       & -6.7  & 1.3 / 1.3 \\
	&                                  &       &           \\
	\hline
	\multirow{9}{*}{A7, A25, D6, D12, D18, D24} & 150, 130, -48.6, -131.3, 90, -90 & +4.2  & 2.0 / 2.1 \\
	& 150, 140, -48.6, -131.3, 90, -90 & +2.6  & 2.1 / 2.1 \\
	& 150, 150, -48.6, -131.3, 90, -90 & +1.6  & 2.1 / 2.1 \\
	&                                  &       &           \\
	& 120, 90, -48.6, -131.3, 90, -90  & -31.9 & 1.1 / 1.0 \\
	& 120, 150, -48.6, -131.3, 90, -90 & +0.3  & 2.2 / 2.3 \\
	&                                  &       &           \\
	& 90, 90, -48.6, -131.3, 90, -90   & -207  & 1.1 / 1.1 \\
	& 90, 150, -48.6, -131.3, 90, -90  & -4.6  & 2.1 / 2.1 \\
	&                                  &       &           \\
	\hline
	\multicolumn{4}{l}{A7 = 120.2; A25 = 121.7; D6 = 48.6; D12 = -131.3; D18 = 129.4; D24 = -50.4; D26 = -50.5} 
\end{longtable}

\subsection{2.4. - Spin density and geometry of the neutral and reduced molecule}

We have also investigated the effect of charging on the spin and geometry of the gas-phase molecule. 

\begin{figure}[h!]
	\includegraphics[width=0.8\textwidth]{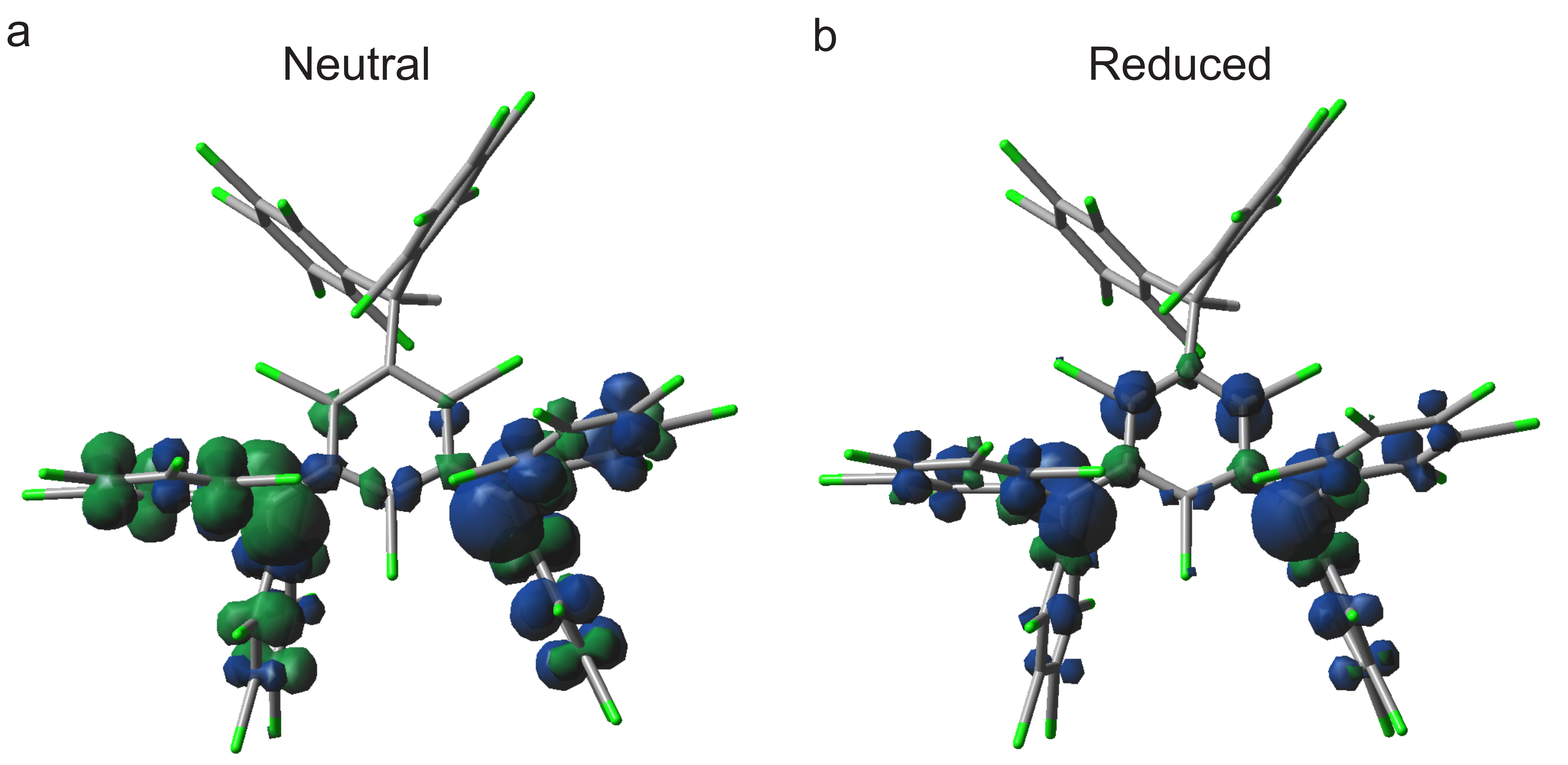}
	\caption{Spin density of the gas-phase neutral and reduced molecule at a torsion angle of 75$^\circ$. According to this calculation, the added charge is located on the central phenyl ring.}
	\label{fig:SD6}
\end{figure}

Fixing the torsion angle to $\theta = 75^\circ$, we calculate the spin densities of the neutral and reduced diradical. At these torsion angle, the former is in the singlet ground state (see \Figure{fig:SD2}) while the latter in the doublet ground state. The result for the neutral molecule is shown in \Figure{fig:SD6}(a): one radical centre presents a majority of $\alpha$-density (blue) and the other a majority of $\beta$-density (green) with both densities localised on their respective pairs of external phenyl rings.
The result for the reduced form at the neutral geometry is shown in \Figure{fig:SD6}(b). Here, both radical centres present the same density, and the inner phenyl ring presents a comparatively higher density than in the neutral case. This can be understood as a consequence of the spin-alternation rule \footnote{A consequence of Lieb theorem for non-alternating lattices according to which the configuration of lower energy would present an alternation of spins such that each centre with $\alpha$-density is surrounded by centres of $\beta$-density, and vice versa} in the inner ring. According to these results, one would expect the extra electron to be located in the inner phenyl ring.

\begin{figure}[h!]
	\includegraphics[width=0.4\textwidth]{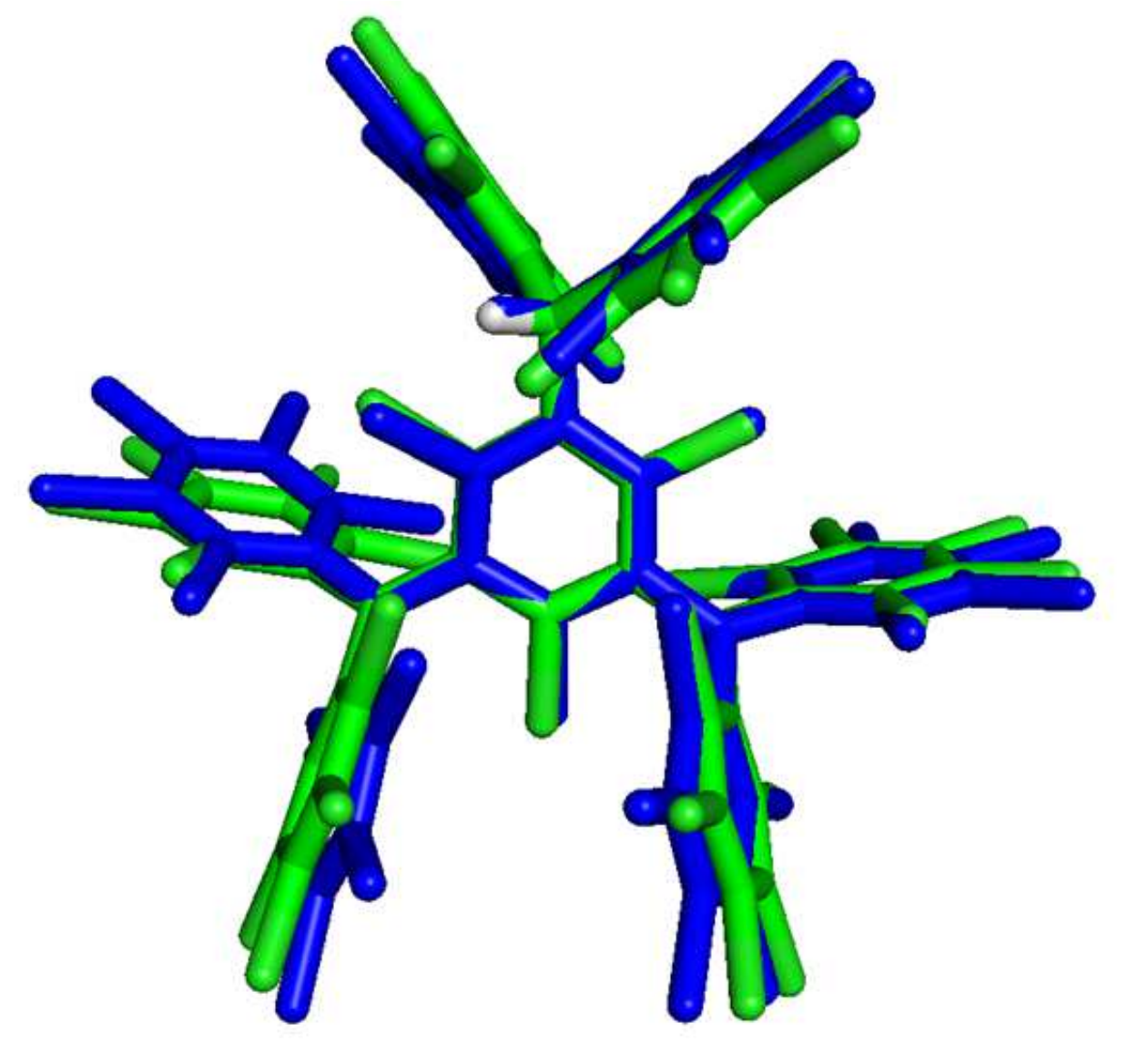}
	\caption{Structural comparison of optimized neutral (green) and anionic (blue) forms, as predicted by B3LYP.}
	\label{fig:SD5}
\end{figure}

In order to compare the geometry of the neutral and reduced diradical, we proceed by characterizing the stationary minima of both forms in gas phase. \Figure{fig:SD5} presents the overlapped structures after imposing one into another by ensuring that the rotation minimizes the RMSD value, following the algorithm proposed by Kabsch\cite{Kabsch1976,Kabsch1978}, and implemented by Kroman and Bratholm\cite{Kroman}. The optimization of the reduced diradical (three spins) has been performed following the same method used to optimize the neutral diradical. An attempt to improve the basis set to include diffuse functions turned out not feasible due to memory problems($\approx$ 1700 basis functions).

The obtained Kabsch RMSD value is 0.674 $\AA$ which indicates that there is a significant difference between both structures. This is an indication that depending on how the molecule sits between the electrodes, it might favor the charging process just because the geometry is more similar to the one that the anion would adopt in the gas phase.

\section{Notes}
The authors declare no competing financial interest.

\bibliography{diradicals_arXive}

\end{document}